\documentclass[11pt,fullpage]{article}

\newcommand{\ol}{\setlength{\itemsep}{0pt.}\begin{enumerate}}
\newcommand{\eol}{\end{enumerate}\setlength{\itemsep}{-\parsep}}
\newcommand{\ignore}[1]{}
\def\C{{\bf C}}
\def\R{{\bf R}}

\title{\bf A polynomial time algorithm to
approximate the mixed volume within a simply
exponential factor}
 
\author{Leonid Gurvits \thanks{%
{\tt gurvits@lanl.gov}. Los Alamos National Laboratory, 
Los Alamos, NM. } 
}

\begin{document}


\begin{titlepage}

\maketitle

\begin{abstract}
Let ${\bf K} = (K_1, \ldots, K_n)$ be an $n$-tuple of convex compact subsets in the Euclidean space $\R^n$, and
let $V(\cdot)$ be the Euclidean volume in $\R^n$. The Minkowski polynomial $V_{{\bf K}}$ is defined
as $V_{{\bf K}}(\lambda_1, \ldots ,\lambda_n) = V(\lambda_1 K_1 +, \cdots , + \lambda_n K_n)$ and the mixed volume $V(K_1, \ldots, K_n)$ as
$$
V(K_1, \ldots, K_n) = \frac{\partial^n}{\partial \lambda_1...\partial
\lambda_n} V_{{\bf K}}(\lambda_1 K_1 +, \cdots, +
\lambda_n K_n).
$$ 
Our main result is a poly-time algorithm which approximates
$V(K_1, \ldots, K_n)$ with multiplicative error $e^n$ and with better rates
if the affine dimensions of most of the sets $K_i$ are small.
Our approach is based on a particular approximation of
$\log(V(K_1, \ldots, K_n))$ by a solution of some convex minimization problem.  We prove the
mixed volume analogues of the Van der Waerden and 
Schrijver-Valiant conjectures on the permanent. These results,
interesting on their own, allow us to justify the abovementioned
approximation by a convex minimization, which is solved using the ellipsoid method and a
randomized poly-time time algorithm for the approximation of the
volume of a convex set.

\end{abstract} 
\end{titlepage}
\newpage

 
\newtheorem{THEOREM}{Theorem}[section]
\newenvironment{theorem}{\begin{THEOREM} \hspace{-.85em} {\bf :} 
}%
                        {\end{THEOREM}}
\newtheorem{LEMMA}[THEOREM]{Lemma}
\newenvironment{lemma}{\begin{LEMMA} \hspace{-.85em} {\bf :} }%
                      {\end{LEMMA}}
\newtheorem{COROLLARY}[THEOREM]{Corollary}
\newenvironment{corollary}{\begin{COROLLARY} \hspace{-.85em} {\bf 
:} }%
                          {\end{COROLLARY}}
\newtheorem{PROPOSITION}[THEOREM]{Proposition}
\newenvironment{proposition}{\begin{PROPOSITION} \hspace{-.85em} 
{\bf :} }%
                            {\end{PROPOSITION}}
\newtheorem{DEFINITION}[THEOREM]{Definition}
\newenvironment{definition}{\begin{DEFINITION} \hspace{-.85em} {\bf 
:} \rm}%
                            {\end{DEFINITION}}
\newtheorem{EXAMPLE}[THEOREM]{Example}
\newenvironment{example}{\begin{EXAMPLE} \hspace{-.85em} {\bf :} 
\rm}%
                            {\end{EXAMPLE}}
\newtheorem{CONJECTURE}[THEOREM]{Conjecture}
\newenvironment{conjecture}{\begin{CONJECTURE} \hspace{-.85em} 
{\bf :} \rm}%
                            {\end{CONJECTURE}}
\newtheorem{PROBLEM}[THEOREM]{Problem}
\newenvironment{problem}{\begin{PROBLEM} \hspace{-.85em} {\bf :} 
\rm}%
                            {\end{PROBLEM}}
\newtheorem{QUESTION}[THEOREM]{Question}
\newenvironment{question}{\begin{QUESTION} \hspace{-.85em} {\bf :} 
\rm}%
                            {\end{QUESTION}}
\newtheorem{REMARK}[THEOREM]{Remark}
\newenvironment{remark}{\begin{REMARK} \hspace{-.85em} {\bf :} 
\rm}%
                            {\end{REMARK}}
\newtheorem{FACT}[THEOREM]{Fact}
\newenvironment{fact}{\begin{FACT} \hspace{-.85em} {\bf :} 
\rm}%
		            {\end{FACT}}

 
\newcommand{\thm}{\begin{theorem}}
\newcommand{\lem}{\begin{lemma}}
\newcommand{\pro}{\begin{proposition}}
\newcommand{\dfn}{\begin{definition}}
\newcommand{\rem}{\begin{remark}}
\newcommand{\xam}{\begin{example}}
\newcommand{\cnj}{\begin{conjecture}}
\newcommand{\prb}{\begin{problem}}
\newcommand{\que}{\begin{question}}
\newcommand{\cor}{\begin{corollary}}
\newcommand{\fac}{\begin{fact}}

\newcommand{\prf}{\noindent{\bf Proof:} }
\newcommand{\ethm}{\end{theorem}}
\newcommand{\elem}{\end{lemma}}
\newcommand{\epro}{\end{proposition}}
\newcommand{\edfn}{\bbox\end{definition}}
\newcommand{\erem}{\bbox\end{remark}}
\newcommand{\exam}{\bbox\end{example}}
\newcommand{\ecnj}{\bbox\end{conjecture}}
\newcommand{\eprb}{\bbox\end{problem}}
\newcommand{\eque}{\bbox\end{question}}
\newcommand{\ecor}{\end{corollary}}
\newcommand{\efac}{\end{fact}}
\newcommand{\eprf}{\bbox}
\newcommand{\beqn}{\begin{equation}}
\newcommand{\eeqn}{\end{equation}}
\newcommand{\wbox}{\mbox{$\sqcap$\llap{$\sqcup$}}}
\newcommand{\bbox}{\vrule height7pt width4pt depth1pt}
\newcommand{\qed}{\bbox}

\newcommand{\rarrow}{\rightarrow}
\newcommand{\larrow}{\leftarrow}
\newcommand{\grad}{\bigtriangledown}

\overfullrule=0pt
\def\setof#1{\lbrace #1 \rbrace}
\section{Introduction}
Let ${\bf K} = (K_1, \ldots, K_n)$ be a $n$-tuple of convex compact subsets in the Euclidean space $\R^n$, and
let $V(\cdot)$ be the Euclidean volume in $R^n$. It 
is a well known result of Hermann Minkowski 
(see for instance \cite{BZ88}), that the value of   
$V_{{\bf K}}(\lambda_1 K_1 + \cdots
\lambda_n K_n)$, 
where $''+''$
denotes Minkowski sum, and $\lambda K$ denotes the dilatation of
$K$ with coefficient $\lambda$, 
is a homogeneous polynomial of degree $n$
in
nonnegative variables $\lambda_1...\lambda_n$ (called the Minkowski polynomial).  
The coefficient $V(K_1, \ldots,K_n)$ of
$\lambda_1\cdot \lambda_2\ldots \cdot \lambda_n$ is called the {\it mixed
volume} of $K_1...K_n$. Alternatively,
$$
V(K_1, \ldots,K_n) = \frac{\partial^n}{\partial \lambda_1...\partial
\lambda_n} V_{{\bf K}}(\lambda_1 K_1 + \cdots +
\lambda_n K_n).
$$
Mixed volume is known to be monotone \cite{BZ88}, namely $K_i \subseteq
L_i$, for $i=1,...,n$, implies $V(K_1, \ldots, K_n) \le V(L_1...L_n)$. In
particular, it is always nonnegative and therefore all the coefficients
of the Minkowski polynomial $V_{{\bf K}}$ are nonnegative real numbers.

The corresponding {\bf Brunn-Minkowski theory}, which is the backbone of convex geometry and its numerous applications,
is about various implications of the fact that the functional $(V_{{\bf K}}(\lambda_1 K_1 + \cdots
\lambda_n K_n))^{\frac{1}{n}}$ is concave on the nonnegative orthant $R^{n}_{+} = \{(\lambda_1,...\lambda_n): \lambda_i \geq 0\}$.
Its generalization, {\bf Alexandrov-Fenchel theory}, is based on the fact that the functionals
$ (\frac{\partial^k}{\partial \lambda_1...\partial \lambda_k} V_{{\bf K}}(0,..,0,\lambda_{k+1},...,\lambda_{n})^{\frac{1}{n-k}}$
are concave on $R^{n-k}_{+}$ for all $1 \leq k \leq n-1$.

The problem of computing the mixed volume of convex bodies is also
important for combinatorics and  algebraic geometry
\cite{da:gr:hu}.  For instance, the number of
toric solutions to a system of $n$ polynomial equations on
${\C}^n$ is upper bounded by---and for a generic system, equal 
to---the mixed volume of the Newton polytopes of
the the corresponding polynomials.  This remarkable result, called the {\bf BKK Theorem},
is covered, for instance, in \cite{Sturm} and \cite{BZ88}). 

\subsection{Previous Work}
The BKK Theorem created an ``industry'' of computing (exactly) the mixed volume of
integer polytopes and its various generalizations; most  algorithms in the area are
of exponential runing time (\cite{huber},\cite{emiris}, \cite{gao} and many more). Most  researchers in the ``industry" don't bother
to formally write down the complexity, rather they describe the actual amount of the computer time. 
Although there was a substantial algorithmic activity on the mixed volume of polytopes prior to \cite{da:gr:hu},
the paper \cite{da:gr:hu} was the first, to our knowledge, systematic complexity-theoretic study in the area.
It followed naturally upon the famous FPRAS algorithms \cite{da:fr} for volumes of convex bodies, solved several natural problems
and posed many important hard questions.  The existence of FPRAS for the mixed volume even for polytopes or ellipsoids is still an open problem.

Efficient polynomial-time probabilistic algorithms that
approximate the mixed volume extremely tightly
(within a (1+$\epsilon$) factor) were developed for some classes of 
well-presented convex bodies \cite{da:gr:hu}. The algorithms
in \cite{da:gr:hu} are based on the multivariate polynomial
interpolation and work if and only if the number $k$
of distinct convex sets in the tuple ${\bf K}$ is ``small", i.e. $k = O(\log(n))$.\\ 
The first efficient probabilistic algorithm 
that provides a $n^{O(n)}$-factor approximation for {\it arbitrary 
well-presented proper
convex bodies} was obtained by 
Barvinok \cite{Bar1}. Barvinok's algorithms start by replacing convex bodies
with ellipsoids. This first step already gives $n^{O(n)}$-factor in the worst case. After that the mixed volume
of ellipsoids is approximated with a simply exponential factor $c^n$ by two randomized algorithms; 
one of those deals with approximation of the mixed discriminant. 

The question of existence of an efficient {\it deterministic}
algorithm for 
approximating the mixed volume of arbitrary 
well-presented proper convex bodies with an error depending only on
the dimension was posed in \cite{da:gr:hu}. The authors quote a lower bound ({\bf B\'ar\'any and F\"uredi bound})
\cite{ba:fu} 
of $\left(\Omega\left(\frac{n}{\log n}\right)\right)^{\frac{n}{2}}$
for the approximation factor of such an algorithm. (Notice that Barvinok's {\it randomized} algorithm
\cite{Bar1} does not beat the {\bf B\'ar\'any and F\"uredi bound}.)

Deterministic polynomial-time algorithms that
approximate the mixed volume with a factor of $n^{O(n)}$ were given,
for a fixed number of distinct proper convex bodies in ${\bf K} = (K_1, \ldots, K_n)$, in \cite{Bar1},
\cite{da:gr:hu}. Finally, a deterministic polynomial-time algorithm that
approximates the mixed volume with a factor of $n^{O(n)}$ in the general case of well-presented
compact convex sets
was given in \cite{GS},\cite{GS1}. 
Let ${\bf A} =(A_1,...,A_n)$ be an $n$-tuple of $n \times n$ complex matrices; the corresponding
determinantal polynomial is defined as $Det_{{\bf A}}(\lambda_1,...,
\lambda_n) = \det(\sum_{1 \leq i \leq n} \lambda_i A_i)$. The mixed discriminant is defined as
$D(A_1,...,A_n) = \frac{\partial^n}{\partial \lambda_1...\partial
\lambda_n}Det_{{\bf A}}( \lambda_1,...,\lambda_n)$.
Similarly to the randomized algorithm from \cite{Bar1}, the algorithm in \cite{GS},\cite{GS1} 
reduced the approximation of the mixed volume of well-presented
compact convex sets to the approximation of the mixed volume of ellipsoids;
this first step gives an $n^{O(n)}$ factor in the
worst case.  And the mixed volume of ellipsoids is
approximated by
$(D(A_1,...,A_n))^{\frac{1}{2}}$ of the corresponding positive semidefinite matrices $A_i \succeq 0$. This second
step adds $\sqrt{3}^{n}$ to the multiplicative approximation error (see inequality (\ref{det_vol}) below).
 
The approximation of the mixed discriminant has also been relaxed
to a convex optimization problem (geometric programming). In order
to prove the accuracy of the convex relaxation, the author proved in \cite{gur} the mixed discriminant analogue
of the Van der Waerden conjecture on permanents of doubly stochastic matrices \cite{minc}, which was posed
by R. V. Bapat in \cite{bapat}.  

To summarize, the interpolational approach from \cite{da:gr:hu} is limited by the restriction that the number
of distinct convex sets is $O(\log(n))$; previous approaches \cite{Bar1}, \cite{GS}, \cite{GS1} can't
give the simply exponential approximation factor $c^n$ because of the initial approximation
of convex sets by ellipsoids.
\subsection{Our Approach}
Assume, modulo deterministic poly-time preprocessing \cite{da:gr:hu}, that the mixed volume $V(K_1, \ldots, K_n) > 0$.
We define the {\bf capacity} of the volume polynomial $ V_{{\bf K}}$ as
\beqn \label{fir-cap}
Cap(V_{{\bf K}}) = \inf_{x_i > 0: 1 \leq i \leq n} \frac{V_{{\bf K}}(x_1,...,x_n)}{\prod_{1 \leq i \leq n} x_i }
\eeqn
Since the coefficients of the volume polynomial $ V_{{\bf K}}$ are nonnegative real numbers we get the upper bound
$$
V(K_1, \ldots, K_n) \leq Cap(V_{{\bf K}}).
$$
The trick is that $\log(Cap(V_{{\bf K}}))$ is a solution
of the following convex minimization problem 
\beqn \label{convex}
\log(Cap(V_{{\bf K}})) = \inf_{y_1+...+y_n = 0} \log(V_{{\bf K}}(e^{y_1},...,e^{y_n})).
\eeqn
{\it Recall that the functional $\log(p( e^{y_1},...,e^{y_n}))$ is convex on $R^n$ if 
$p(x_1,...,x_n)$ is any polynomial with nonnegative coefficients.
More generally, a sum of log-convex functionals is log-convex.}

We view $Cap(V_{{\bf K}})$ as an approximation for the mixed volume $V(K_1, \ldots, K_n)$.  To justify this
we prove the lower bound 
$$
V(K_1, \ldots, K_n) \geq \frac{n!}{n^n} Cap(V_{{\bf K}}) \approx e^{-n} Cap(V_{{\bf K}}),
$$
which is the mixed volume analogue
of the Van der Waerden conjecture on the permanent. We also present better upper bounds when ``most" of the convex sets $K_i$ have small affine dimension,
which are analogues of the Schrijver-Valiant conjecture, posed in \cite{schr-val} and proved in \cite{schr}.

The idea of our approach is very similar to our treatment of {\bf H-Stable} polynomials in \cite{ejc-2008}.
Recall that a homogeneous polynomial $p(x_1,...,x_n)$ with nonnegative coefficients is called {\bf H-Stable}
if $p(z_1,...,z_n) \neq 0$ provided that the real parts $Re(z_i) > 0, 1 \leq i \leq n$.

Not all Minkowski polynomials $V_{{\bf K}}$ are H-Stable: any univariate polynomial with nonnegative coefficients
$S(x) = \sum_{0 \leq i \leq n} {n \choose i} a_i x^i$ such that $ a_i^2 \geq a_{i-1} a_{i-1}, 1 \leq i \leq n-1$ can be presented
as $S(x) = V(A + xB)$ for some convex compact subsets (simplexes) $A,B \subset R^n$ \cite{sheppard}.
Fortunately, a modification of the inductive proof in \cite{ejc-2008} works for Minkowski polynomials and presented in the next Section.

After establishing the mixed volume analogues of the Van der Waerden and  the Schrijver-Valiant (permanental) conjectures, 
we present a randomized poly-time algorithm to solve the problem (\ref{convex}) based on the ellipsoid method
and randomized poly-time algorithms for the volume approximation.   
Together with the mixed volume analogues of the Van der Waerden conjecture this gives a randomized poly-time algorithm to approximate the mixed volume $V(K_1, \ldots, K_n)$ within
relative accuracy $e^n$.
Notice that, in view of the {\bf B\'ar\'any and F\"uredi bound}, this cannot be achieved by a deterministic poly-time oracle
algorithm.
We use the ellipsoid method because of its robustness: we deal essentially with a random oracle which
computes $\log \left(V_{{\bf K}}(e^{y_1},...,e^{y_n}) \right)$ with an additive small error $\epsilon$ ;
we use this oracle to get an approximation of the gradient of $\log \left(V_{{\bf K}}(e^{y_1},...,e^{y_n}) \right)$.\\

\section{ Van der Waerden and Schrijver-Valiant conjectures for the mixed volume}
{\it Consult Appendix {\bf A} for the proofs of results in this section.}
\subsection{The mixed volume analogue of the Van der Waerden-Falikman-Egorychev inequality}
\thm \label{VDW}
\begin{enumerate}
\item
Let ${\bf K} = (K_1, \ldots, K_n)$ be a $n$-tuple of convex compact subsets in the Euclidean space $\R^n$.
The mixed volume $V(K_1, \ldots, K_n)$ satisfies the next lower bound:
\beqn \label{waer ineq}
V(K_1, \ldots, K_n) \geq \frac{n!}{n^n} Cap(V_{{\bf K}})
\eeqn
\item
The equality in (\ref{waer ineq}) is attained if and only if either the mixed volume $V(K_1, \ldots, K_n) = 0$
or $K_i = a_i K_1 + b_i : a_i > 0, b_i \in R^n ; 2 \leq i \leq n$.
\end{enumerate}
\ethm
\subsection{The mixed volume analogue of the Schrijver-Valiant conjecture}

\dfn
\begin{enumerate}
\item
Let $n \geq k \geq 1$ be two integers. We define the univariate
polynomial 
$$
sv_{n,k}(x) = 1+\sum_{1 \leq i \leq k} \left(\frac{x}{n} \right)^{i} {n \choose i}.
$$
Note that $sv_{n,n}(x) = \left(1 + \frac{x}{n} \right)^{n}$.  Next we define the
 functions :
\beqn \label{factors}
\lambda(n,k) = \left(\min_{x > 0} (\frac{sv_{n,k}(x)}{x}) \right)^{-1}
\eeqn
\rem \label{old}
It was observed in \cite{stoc-2006} that 
\beqn \label{factor}
\lambda(k,k) = g(k) =: \left(\frac{k-1}{k} \right)^{k-1}, k \geq 1 ; \prod_{1 \leq k \leq n} g(k) = \frac{n!}{n^n}
\eeqn
The following inequalities are easily verified: 
\beqn \label{monoton}
\lambda(n,k) < \lambda(n,l) : n \geq k > l \geq 1; \lambda(m,k) > \lambda(n,k) : n > m \geq k.
\eeqn
It follows that
\beqn \label{exp-trunc}
\lambda(\infty,k) =: \lim_{n \rightarrow \infty}\lambda(n,k) = \left(\min_{t > 0} 
\frac{\sum_{0 \leq i \leq k} \frac{t^i}{i!}}{t} \right)^{-1}
\eeqn
The equality $\lambda(n,2) = \left(1+ \sqrt{2} \sqrt{\frac{n-1}{n}} \right)^{-1} \geq (1+ \sqrt{2})^{-1}$ follows from basic calculus.
\erem
\item
Let $n \geq m \geq 1$ be two integers. An univariate polynomial with nonnegative coefficients $R(t) = \sum_{0 \leq i \leq m} a_i t^i$ is called $n-Newton$ if
it satisfies the following inequalities :
\beqn \label {n-newton}
NIs : \left(\frac{a_i}{{n \choose i}} \right)^2 \geq \frac{a_{i-1}}{{n \choose i-1}} \frac{a_{i+1}}{{n \choose i+1}} : 1 \leq i \leq m-1
\eeqn

(The standard Newton's inequalities correspond to the case $n =m$ are satisfied if all the roots of $p$ are real.)
\end{enumerate}
\edfn

The main mathematical result in this paper is the following theorem:
\thm \label{S-V-G}
Let ${\bf K} = (K_1, \ldots, K_n)$ be a $n$-tuple of convex compact subsets in the Euclidean space $\R^n$ and
$aff(i)$ be the affine dimension of $K_i$, $1 \leq i \leq n$. Then the following inequality holds:
\beqn \label{main result}
Cap(V_{{\bf K}}) \geq V(K_1, \ldots, K_n) \geq \prod_{1 \leq i \leq n} \lambda(i, D(i)) Cap(V_{{\bf K}}); D(i) = \min \left(i, aff(i) \right), 1 \leq i \leq n.
\eeqn
\ethm

\cor \label{Cor}
Suppose that $aff(i) \leq k : k+1 \leq i \leq n$ then
\beqn \label{schr ineq}
V(K_1, \ldots, K_n) \geq \frac{k!}{k^k} \lambda(n,k)^{n-k} Cap(V_{{\bf K}}).
\eeqn
If $k=2$ we get the inequality 
\beqn \label{af=2}
V(K_1, \ldots, K_n) \geq \frac{1}{2}(1 + \sqrt{2})^{2-n}  Cap(V_{{\bf K}})
\eeqn
\ecor

\subsection{Comparison with previous results}
The inequality (\ref{waer ineq}) is an analogue of the famous Van der Waerden conjecture \cite{minc}, proved in \cite{fal} and \cite{ego}, on the permanent of
doubly-stochastic matrices. Indeed, consider the ``boxes" $K_{i} = \{(x_1,...,x_n) : 0 \leq x_j \leq A(i,j), 1 \leq j \leq n\}$.
Then the mixed volume is equal to the permanent, 
$$
V({\bf K}) = V(K_1, \ldots, K_n) = Perm(A),
$$
and if the $n \times n$ matrix $A$ is doubly-stochastic
then $Cap(V_{{\bf K}}) = 1$.

Though this mixed volume representation of the permanent 
has been known since the publication of \cite{ego} if not earlier,
the author is not aware of any attempts prior to the present paper 
to generalize the Van der Waerden conjecture to the mixed volume.  We think
that our version, stated in terms of the {\bf {\bf capacity}}, is most natural and useful.

The inequality (\ref{schr ineq}) is an analogue of Schrijver's lower bound \cite{schr}, \cite{ejc-2008} on
the number of perfect matchings in $k$-regular bipartite graphs: affine dimensions play role of the degrees of vertices.

The reader familiar with \cite{ejc-2008} can recognize the similarity between inequalities (\ref{main result}), (\ref{waer ineq}), (\ref{schr ineq})
and the corresponding inequalities in \cite{ejc-2008}, proved for {\bf H-Stable} polynomials. The method of proof in the present paper
is also similar to the one in \cite{ejc-2008} (in spite of the fact that not all Minkowski polynomials $V_{{\bf K}}$ are {\bf H-Stable}).
But we get worse constants: for instance, if $k=2$, in the notation of (\ref{schr ineq}), then in the {\bf H-Stable} case
we get the factor $2^{-n+1}$ instead of the 
$\frac{1}{2} (1 + \sqrt{2})^{2-n}$ obtained in this paper. 
Whether the latter factor is asymptotically
sharp
is an open problem.

\subsection{The idea of our proof}
Our proof of Bapat's conjecture in \cite{gur}, i.e. of Van der Waerden conjecture for the mixed discriminant, is an adaptation
of Egorychev's proof in \cite{ego}. In contrast, the proofs in \cite{ejc-2008} and in the present paper have practically nothing in common
either with Egorychev's proof or with Falikman's proof in \cite{fal}.\\
\begin{enumerate}
\item 
{\it How do we prove the lower bounds?}\\
We follow the general approach by the present author, introduced in \cite{ejc-2008}.\\
We associate with the Minkowski polynomial $V_{{\bf K}}$ a sequence of polynomials:\\
$$
q_n = V_{{\bf K}}, q_i(x_1,...,x_i) = \frac{\partial^{n-i}}{\partial x_{i+1}...\partial x_n}q_{n}(x_1,...,x_i,0,...,0), 1 \leq i \leq n-1.
$$
Note that $q_{1}(x) = V(K_1, \ldots, K_n) x$. Everything follows from the next inequality
$$
Cap(q_i) \geq \lambda \left(i+1, \min(i+1, aff(i+1)) \right) Cap(q_{i+1}), 1 \leq i \leq n-1.
$$
Not surprisingly, we do use the (still hard to prove) Alexandrov-Fenchel inequalities to prove this crucial inequality.
(In contrast, the {\bf H-Stable}
case in \cite{ejc-2008} required just the elementary AG inequality).

\item
{\it How do we prove the uniqueness?} 

The uniqueness proofs in \cite{ego} and \cite{gur} are critically based on the known characterization of
the equality cases in the Alexandrov inequalities for the mixed discriminant. In the case of the Alexandrov-Fenchel inequalities
for the mixed volume such a characterization is not known. Luckily, the method of our proof of the lower bound (\ref{waer ineq})
allows us to use the well known characterization of  equality in the (much simpler) Brunn-Minkowski inequality. The uniqueness proof in the present
paper is very similar to the uniqueness proof in \cite{ejc-2008}. The fundamental tool in \cite{ejc-2008} was G{\aa}rding's famous (and not hard to prove) 
result on the convexity of the hyperbolic cone.
\end{enumerate}

\section{Convex Optimization Relaxation of the Mixed Volume}
Inequalities (\ref{main result}, \ref{waer ineq}, \ref{schr ineq}) justify the following strategy for approximation of
the mixed volume $V(K_1, \ldots, K_n)$ within a simply exponential multiplicative factor: solve the convex optimization
problem (\ref{convex}) with an additive $O(1)$ error. We follow here the approach from \cite{GS, GS1} which dealt with
the following problem:
\beqn \label{DET}
\log(Cap(Det_{{\bf A}})) = \inf_{y_1+...+y_n = 0} \log \left(\det(\sum_{1 \leq i \leq n} e^{y_i} A_i) \right) : A_i \succeq 0.
\eeqn
The main difference between the two problems is that the value and the gradient of determinantal polynomials
can be exactly evaluated in deterministic polynomial time. The case of the Minkowski polynomials $V_{{\bf K}}$ requires
some extra care. Yet, this is done using standard and well known tools. This section of the paper is fairly routine
and can be easily reproduced by any convex optimization professional.

We now give an overview of the main points:
\subsection{ A brief overview}
\begin{enumerate}
\item {\bf Representations of convex sets and a priori ball for the convex relaxation}: we deal in this paper with two types
of representations. First, similar to \cite{da:gr:hu} we consider {\it well-presented} convex compact sets; second, motivated
by algebraic applications and the {\bf BKK} theorem, we consider integer polytopes given as a list of extreme points. In both cases,
we start with deterministic poly-time preprocessing which transforms the initial tuple ${\bf K} \in R^n$ into a collection
of indecomposable tuples
$$
{\bf K^{(1)}} \in R^{d_1},...,{\bf K^{(i)}} \in R^{d_i}; \sum_{1 \leq j \leq i} d_j = n
$$
such that
the mixed volume $V({\bf K}) = \prod_{1 \leq j \leq i}V({\bf K^{(i)}})$. The tuple ${\bf K}$ is indecomposable if and only if the minimum in (\ref{convex}) is attained and unique.
The preprocessing is essentially the same as in \cite{GS1}.

After this preprocessing we deal only with the indecomposable case and get a priori ball which is guaranteed to contain the unique
minimizer of (\ref{convex}). The radius of this ball is expressed in terms of the complexity of the corresponding
representation:\\
$r \leq O(n^{2}(\log(n) + <{\bf K}>))$, where $<{\bf K}>$ is the complexity of the initial tuple ${\bf K}$.

This part is fairly similar to the analogous problem for (\ref{DET}) treated in \cite{GS1}.
\item {\bf Lipschitz Property and Rounding:} In the course of our algorithm we need to evaluate the volumes
$V(e^{y_1}K_1 + \ldots +e^{y_n}K_n)$ and the mixed volumes $V(K_i,B,...,B), B = e^{y_1}K_1 + \ldots +e^{y_n}K_n$. This requires
a well-presentation of the Minkowski sum $B = e^{y_1}K_1 +...+e^{y_n}K_n$. Given the well-presentation
of ${\bf K}$ one gets a well-presentation of $a_1 K_1 +...+a_n K_n$ if the sizes of positive rational numbers $a_i$
are bounded by $poly(n,<{\bf K}>)$ \cite{da:gr:hu}. Therefore we need a rounding procedure, which requires us
to keep only a ``small" number of fractional bits of $y_i$ (integer bits are taken care of by a priori ball).   This can be done using the Lipschitz property (\ref{Lipsh})
of $\log\left(V(e^{y_1}K_1 + \ldots +e^{y_n}K_n) \right)$ (which is just Euler's identity for homogeneous functionals) and its partial
derivatives, proved in Lemma(\ref{log-con}).

\item {\bf Complexity of our Algorithm}: We give an upper bound on the number of calls to oracles for Minkowski
sums $a_1 K_1 +...+a_n K_n$. The number of calls to oracles for the initial tuple ${\bf K}$ will be larger
but still polynomial (see the discussion in \cite{belkin}  
in the context of surface area computation,
which is, up to a constant, the mixed volume $V(Ball_{n}(1),A,...,A)$). 
\item {\bf Ellipsoid Method with noisy first order oracle}: let $g(.)$ be a differentiable convex functional defined
on the closed ball $Ball_{n}(r) = \{X \in R^n : <X,X> \leq r^2\}$ and $Var(g) = \max_{X \in Ball_{n}(r)}g(X) - \min_ {X \in Ball_{n}(r)}g(X)$.
The standard version of the ellipsoid method requires exact values of the function and its gradient. Fortunately, there exists a noisy version
\cite{YN}, which needs approximations of the value $\overline{g(X)}$ and of the gradient $\overline{(\bigtriangledown g)(X)}$ such that
$$
sup_{Y,X \in Ball_{n}(r)} |(\bar{g}(X) + <\overline{(\bigtriangledown f)}(X),Y-X>) - (g(X) + (\bigtriangledown g)(X),Y-X>)| \leq \delta Var(g).
$$
In our case, $g(y_1,...,y_n) = \log(V_{{\bf K}}(e^{y_1},...,e^{y_n}))$. We get the additive approximation of $g(y_1,...,y_n)$
using a FPRAS for the volume approximation and the additive approximation of $(\bigtriangledown g)(X)$ using FPRAS from \cite{da:gr:hu}
for approximating the ``simple" mixed volume (generalized surface area) $V(K_i,K,...,K), K = \sum_{1 \leq i \leq n}e^{y_i} K_i $.
\end{enumerate}
\subsection{Representations of convex compact sets}
Following \cite{da:gr:hu} we consider the following well-presentation of convex compact set $K_i \subset R^n, 1 \leq i \leq n$:
a weak membership oracle for $K$ together with a rational $n \times n$ matrix $A_i$ and a rational vector $Y_i \in R^n$ such that
\beqn \label{well-pres}
Y_i + A_i(Ball_{n}(1)) \subset K_i \subset Y_i + n\sqrt{n+1} A_i(Ball_{n}(1))
\eeqn  

We define the size $<{\bf K}>$ as the maximum of bit sizes of entries of matrices $A_i,1 \leq i \leq n$.  
Since the mixed volume $V(K_1, \ldots, K_n) = V(K_1 + \{-Y_1\},...,K_n + \{-Y_n \}$,
we will assume WLOG that $Y_i = 0, 1 \leq i \leq n$. This assumption implies 
the following identity for affine dimensions 
\beqn \label{ranks}
aff(\sum_{i \in S} K_i) = Rank(\sum_{i \in S} A_i A_i^{T}), S \subset \{1,...,n\}.
\eeqn
\dfn 
An $n$-tuple ${\bf K} = (K_1, \ldots, K_n)$ of convex compact subsets in $R^n$ is called indecomposable
if $aff(\sum_{i \in S} K_i) > Card(S) : S \subset \{1,...,n\}, 1 \leq Card(S) < n$. \\
We consider, similar to \cite{GS1}, $n(n-1)$ auxiliary $n$-tuples 
${\bf K}^{ij}$, where ${\bf K}^{ij}$ is obtained from ${\bf K}$ by
substituting 
$K_i$ instead of $K_j$. Notice that
\beqn \label{sum in vol}
V(x_1 K_1 +...+x_nK_n) =  x_1 x_2...x_n(V({\bf K}) + \frac{1}{2} \sum_{1 \leq i,j \leq n} \frac{x_i}{x_j} V({\bf K}^{ij})) + ...
\eeqn
\edfn
It follows from (\ref{ranks}) that the $n$-tuple ${\bf K} = (K_1, \ldots, K_n)$ of well-presented convex sets is
indecomposable iff the $n$-tuple of positive semidefinite matrices ${\bf Q} = (Q_1...Q_n) : Q_i = A_i A_i^{T} $ is
fully indecomposable as defined in \cite{GS1}, which implies that indecomposability of ${\bf K}$ is equivalent
to the inequalities $V({\bf K}^{ij}) > 0 : 1 \leq i,j \leq n$. Here $V({\bf K}^{ij})$ stands for the mixed volume of the $n$-tuple ${\bf K}^{ij}$.\\

{\it It was proved in \cite{GS1} that an $n$-tuple of positive semidefinite matrices ${\bf Q} = (Q_1...Q_n)$ is indecomposable
if and only if there exists an unique minimum in the optimization problem 
$$
\inf_{y_1+...+y_n = 0} \log(\det(\sum_{1 \leq i \leq n} e^{y_i} Q_i)).
$$
In the same way, an $n$-tuple ${\bf K} = (K_1, \ldots, K_n)$ of convex compact subsets in $R^n$ is indecomposable
if and only if there exists an unique minimum in the optimization problem (\ref{convex}).}  

Applying the decomposition algorithm from {\it Section 2} in \cite{GS1} to $n$-tuple of positive semidefinite matrices ${\bf Q} = (Q_1...Q_n)$, 
we can, by deterministic poly-time preproprocessing, determine whether or not 
the $n$-tuple of convex compact subsets ${\bf K}$ is indecomposable and if not, 
factor the mixed volume as $V({\bf K}) = \prod_{1 \leq j \leq m \leq n} V({\bf K_j})$.
Here the $n(j)$-tuple ${\bf K_j} = (K_{j,1},...,K_{j,n(j)}) \subset R^{n(j)}$ is well presented and indecomposable,
$\sum_{1 \leq j \leq m} n(j) = n$ and the sizes $<{\bf K_j}> ~\leq~ <{\bf K}> + poly(n)$.\\

Based on the above remarks, we will deal from now on only with indecomposable well-presented tuples of convex compact sets.
Moreover, to simplify the exposition, we assume WLOG that the matrices $A_i$ in (\ref{well-pres}) are integer.\\

Let ${\cal E}_{A}$ be the ellipsoid $A(Ball_{n}(1))$ in $R^n$. The following inequality, proved in \cite{Bar1}, connects the mixed
volume of ellipsoids and the corresponding mixed discriminant:
\beqn
\label{det_vol}
3^{-\frac{n+1}{2}} v_n D^{\frac{1}{2}}(A_1 (A_1)^{T},...,A_n (A_n)^{T}) \leq 
V({\cal E}_{A_1}...{\cal E}_{A_n}) \leq v_n
D^{\frac{1}{2}}(A_1 (A_1)^{T},...,A_n (A_n)^{T}).
\eeqn
Here $v_n$ is the volume of the unit ball in $\R^n$. 
\subsection{Properties of volume polynomials: Lipschitz, bound on the second derivative, a priori ball}
\pro \label{property}
\begin{enumerate}
\item {\bf Lipschitz Property}.\\
Let $p(x_1,...,x_n)$ be a nonzero homogeneous polynomial of degree $n$ with nonnegative coefficients, $x_i = e^{y_i}$.
Then
\beqn
\frac{\partial}{\partial y_i} \log(p(e^{y_1},...,e^{y_n})) = \frac{\frac{\partial}{\partial x_i}p(x_1,...,x_n) e^{y_i}}{p(x_1,...,x_n)}
\eeqn
It follows from the Euler's identity that $\sum_{1 \leq i \leq n} \frac{\partial}{\partial y_i} \log(p(e^{y_1},...,e^{y_n})) = n$,
therefore the functional $f(y_1,...,y_n) = \log(p(e^{y_1},...,e^{y_n}))$ is Lipschitz on $R^n$:
\beqn \label{Lipsh}
|f(y_1 + \delta_1,...,y_n + \delta_n) - f(y_1,...,y_n)| \leq n ||\Delta||_{\infty} \leq  n ||\Delta||_{2}
\eeqn
\item {\bf Upper bound on second derivatives}.\\
Let us fix real numbers  $y_1,...,y_{i-1},y_{i+1},...,y_n$ and define univariate function 
$q(y_i) = \log(V_{\bf K}(e^{y_1},...,e^{y_i},...,e^{y_n}))$.
Notice that $e^{q(y_i)} = \sum_{0 \leq j \leq aff(K_i)} a_j e^{jy}, a_j \geq 0$. 
\pro \label{log-con}
\beqn \label{sec-der}
0 \leq q^{''}(y) \leq aff(K_i)\;.
\eeqn 
({\it Lemma(\ref{vtor}) in Appendix {\bf B} proves a more general inequality.})
\epro 
\item {\bf A Priori Ball result from \cite{GS1}}.\\
Let $p \in Hom_{+}(n,n)$, $p(x_1,...,x_n) = x_1 x_2...x_n(a + \frac{1}{2} \sum_{1 \leq i \neq j \leq n} b^{i,j}\frac{x_i}{x_j}) +...$.\\
Assume that $min_{1 \leq i \neq j \leq n} b^{i,j} =: Stf(p) > 0$. Then there exists an unique minimizer $(z_1,...,z_n),\sum_{1 \leq i \leq n}z_i = 0 $ such that
$$
\log(p(z_1,...,z_n)) = \min_{\sum_{1 \leq i \leq n}y_i = 0} \left(\log(p(e^{y_{1}},...,e^{y_{n}}) \right) . 
$$
Moreover,
\beqn \label{ball}
|z_i - z_j| \leq \log \left(\frac{2 Cap(p)}{Stf(p)} \right)
\eeqn
\end{enumerate}
\epro
The next proposition directly adapts Lemma 4.1 from \cite{GS1} to the mixed volume situation, using  Barvinok's
inequality (\ref{det_vol}).
\pro
Consider an indecomposable $n$-tuple of convex compact sets ${\bf K} = (K_1, \ldots, K_n)$ with the well-presentation
$A_i(Ball_{n}(1)) \subset K_i \subset y + n\sqrt{n+1} A_i(Ball_{n}(1)), 1 \leq i \leq n$ with integer $n \times n$ matrices $A_i$.
Then the minimum in the convex optimization problem (\ref{convex}) is attained and unique. The unique
minimizing vector 
$(z_1,...,z_n), \sum_{1 \leq i \leq n} z_i = 0$ satisfies the following inequalities: 
\beqn \label{mixvol ball}
|z_i - z_j| \leq O(n^{\frac{3}{2}}(\log(n) + <{\bf K}>)) ; ||z_1,...,z_n)||_2 \leq O(n^{2}(\log(n) + <{\bf K}>))
\eeqn
In other words the convex optimization problem (\ref{convex}) can be solved on the following ball in $R^{n-1}$:
\beqn \label{ball-rest}
Apr({\bf K}) = \{(z_1,...,z_n) :||z_1,...,z_n)||_2 \leq O(n^{2}(\log(n) + <{\bf K}>)),\sum_{1 \leq i \leq n} z_i = 0 \}
\eeqn 
The following inequality follows from the Lipschitz property (\ref{Lipsh}):
\beqn \label{bound-lip}
|\log(V_{{\bf K}}(e^{y_1},...,e^{y_n}) - \log(V_{{\bf K}}(e^{l_1},...,e^{l_n})| \leq O(n^{3}(\log(n) + <{\bf K}>)) : Y,L \in Apr({\bf K}).
\eeqn
\epro
\subsection{Ellipsoid method with noisy first order oracles}
We recall the following fundamental result \cite{YN}:\\

{\it Let $f(Y)$ be differentiable convex functional defined on the ball $Ball_{n}(r) = \{Y \in R^n : <Y,Y> \leq r^2 \}$
of radius $r$. Let $Var(f) = \max_{Y \in Ball_{n}(r)}f(Y) - \min_{Y \in Ball_{n}(r)}f(Y)$.
Assume that at each vector $Y \in Ball_{n}(r)$ we have an oracle evaluating a value $g(Y)$ such that $|g(Y) - f(Y)| \leq 0.2 \delta Var(f)$
and the vector $gr(Y) \in R^n$ such that $||gr(Y) - (\bigtriangledown f)(Y)||_{2} \leq 0.2 \delta r^{-1}Var(f)$ ( here $(\bigtriangledown f)(Y)$
is the gradient of $f$ evaluated at $Y$).  Then the {\bf Ellipsoid method} finds a vector $Z \in Ball_{n}(r)$
such that $f(Z) \leq \min_{Y \in Ball_{n}(r)}f(Y) + \epsilon Var(f), \epsilon > \delta$. The method requires
$O(n^2 \log (\frac{1}{\epsilon -\delta}))$ oracle calls
plus $O(n^2)$ elementary operations to run the algorithm itself. }
\subsection{Putting things together}
Here we take advantage of randomized algorithms which can evaluate $\log(Vol(K))$,for a well-presented convex set $K$, with an additive error $\epsilon$
and failure probability $\delta$ in $O(\epsilon^{-k} n^l \log(\frac{1}{\delta}))$ oracle calls.  For instance, the best current
algorithm \cite{vemp} gives $k=2, l =4$. We will need below to evaluate volumes $V(\sum_{1 \leq i \leq n} x_i K_i)$.
In our case the functional $f = \log(V_{{\bf K}}(e^{y_1},...,e^{y_n})$ defined on ball $Apr({\bf K})$ of radius
$O(n^{2}(\log(n) + <{\bf K}>))$ with the variance $Var(f) \leq O(n^{3}(\log(n) + <{\bf K}>))$.
Theorem \ref{VDW} gives the bound:
\beqn 
\log(V({\bf K})) \leq (\min_{Y \in Apr({\bf K})} f(Y)) \leq \log(V({\bf K})) + \log(\frac{n^n}{n!}) \approx \log(V({\bf K})) + n.
\eeqn
Therefore, to approximate the mixed volume  $V({\bf K}$) up to a multiplicative factor 
of $e^n$ it is sufficient to find 
$Z \in Apr({\bf K})$ such that $f(Z) \leq \min_{Y \in Apr({\bf K})} f(Y) + O(1)$. In order to get that via the Ellipsoid
method we need to approximate $\log(V_{{\bf K}}(e^{y_1},...,e^{y_n}))$ with the additive error
$O(Var(f)^{-1}) = O(n^{-3}(\log(n) + <{\bf K}>)^{-1} )$ and its gradient with the additive $l_2$ error 
$O(n^{-2}(\log(n) + <{\bf K}>)^{-1})$.\\
\begin{enumerate}
\item {\bf Approximation of $\log(V_{{\bf K}}(e^{y_1},...,e^{y_n}))$ with failure probability $\delta$}.
The complexity is $O(n^{10} (\log(n) + <{\bf K}>)^{2} \log(\delta^{-1}))$
\item {\bf Approximation of the partial derivatives}.
Let $x_i = e^{y_i}$ and recall that the partial derivatives are
$$
\beta_i = \frac{\partial}{\partial y_i} \log(V_{{\bf K}}(e^{y_1},...,e^{y_n})) = 
\frac{\frac{\partial}{\partial x_i}V_{{\bf K}}(x_1,...,x_n) e^{y_i}} {V_{{\bf K}}(x_1,...,x_n)}\;.
$$
Suppose that $ 0 \leq 1-a \leq \frac{\gamma_i}{\beta_i} \leq 1+a$. It follows from the Euler's identity
that $ \sum_{1 \leq i \leq n} |\gamma_i - \beta_i| \leq a$. If $a = O(n^{-2}(\log(n) + <{\bf K}>)^{-1})$,
then the vector $(\gamma_1,...,\gamma_n)$ is the needed approximation of the gradient.

Notice that $ \Gamma_i = \frac{\partial}{\partial x_i}V_{{\bf K}}(x_1,...,x_n) = \frac{1}{(n-1)!}V(A,B,...,B)$,
where the convex sets $A = K_i$ and $B = \sum_{1 \leq i \leq n}e^{y_j} K_j$. The randomized algorithm from  
\cite{da:gr:hu} approximates $V(A,B,...,B)$ with the complexity $O(n^{4 + o(1)} \epsilon^{-(2+o(1)} \log(\delta)$.
This gives the needed approximation of the gradient with the complexity $n O(n^{8 + o(1)}(\log(n) + <{\bf K}>)^{2 +o(1)} \log(\delta^{-1}))$.
\item {\bf Controlling the failure probability $\delta$ }.
We need to approximate $O(n^2 \log(Var(f)))$ values and gradients. To achieve a probability of success $\frac{3}{4}$ we
need that
$$ 
\delta \approx \frac{1}{4} \left(n^2 (n^{\frac{5}{2}}(\log(n) + <{\bf K}>)) \right)^{-1}. 
$$
This gives
$\log((\delta)^{-1}) \approx O(\log(n) + \log(\log(n) + <{\bf K}>)).$
\end{enumerate}

\rem
Let $g(y_1) = \log(V_{{\bf K}}(e^{y_1},...,e^{y_n}))$ and $\overline{g(y)} = g(y) + h(y), |h(y)| \leq a$. We present here an alternative elementary way to approximate the partial
derivative $g^{\prime}(y_1)$.

Recall that
the function $g(.)$ is convex and $0 \leq g^{\prime \prime}(x) \leq n : x \in R$.
It follows that 
\beqn \label{delta}
|\frac{\overline{g(y + \delta)} - \overline{g(y)}}{\delta} - g^{\prime}(y)| \leq \frac{n}{2} \delta + \frac{2a}{\delta}
\eeqn
The optimal value in (\ref{delta}), $\delta_{opt} = 2 \sqrt{\frac{a}{n}}$, gives the bound
\beqn \label{opt-delta}
|\frac{\overline{g(y + \delta_{opt})} - \overline{g(y)}}{\delta_{opt}} - g^{\prime}(y)| \leq 2 \sqrt{na}.
\eeqn
The simple ``estimator" (\ref{delta}) can be used instead of the interpolational algorithm from \cite{da:gr:hu}, but
its worst-case complexity seems to be higher than that from \cite{da:gr:hu}.
\erem

\thm \label{Ellip}
Given a $n$-tuple ${\bf K}$ of well-presented convex compact sets in $R^n$ there is a poly-time algorithm which
computes the number $AV({\bf K})$ such that
$$
Prob \{ 1 \leq \frac{AV({\bf K})}{V({\bf K})} \leq 2 \prod_{1 \leq i \leq n} \lambda(i, \min(i, aff(i))) \leq 2 \frac{n^n}{n!} \} \geq.75 
$$
The complexity of the algorithm, neglecting the $\log$ terms, is bounded by \\ 
$O(n^{12} (\log(n) + <{\bf K}>)^{2})$.
\ethm 
Next, we focus
on the case of  Newton polytopes, in other words, polytopes with integer vertices. 
{\em I.e}. we will consider the mixed volumes
$V({\bf P}) = V(P_1,...,P_n)$, where 
$$
P_i = Hull \left( \{v_{i,j} : 1 \leq j \leq m(i), v_{i,j} \in Z_{+}^{n} \} \right).
$$

We define 
$$
d(i) = \min\{k :P_i \subset k Hull(0,e_1,...,e_n)\},
$$
i.e. $d(i)$ is the maximum coordinate
sum attained on $P_i$. It follows from the monotonicity of the mixed volume that $V(P_1,...,P_n) \leq \prod_{1 \leq i \leq k} d(i)$.
Such polytopes are well-presented if, for instance, they are given as a list of $poly(n)$ vertices. This
case corresponds to a system of sparse polynomial equations. Notice that the value $V(P_1,...,P_n)$ is either zero or an integer  
({\bf BKK} Theorem) and the {\bf capacity} $Cap(V_{{\bf P}}) \leq \frac{n^n}{n!}\prod_{1 \leq i \leq k} d(i)$ (inequality (\ref{waer ineq})).

The next theorem is proved in the same way
as Theorem (\ref{Ellip}).
\thm \label{polytope}
Given $n$-tuple of ${\bf P} =(P_1,...,P_n)$ of well-presented integer polytopes in $R^n$ there is a poly-time algorithm which
computes the number $AV({\bf P})$ such that
$$
Prob \{ 1 \leq \frac{AV({\bf P})}{V({\bf P})} \leq 2 \prod_{1 \leq i \leq n} \lambda(i, \min(i, aff(P_{i}))) \leq 2 \frac{n^n}{n!} \} \geq.75 
$$ 
The complexity of the algorithm, neglecting the $\log$ terms, is bounded by $O(n^9(n + \log(\prod_{1 \leq i \leq n} d_i))^2)$.
\ethm

\section{Open Problems}
\begin{enumerate}
\item Prove that for ``random" convex sets $\frac{V(K_1, \ldots, K_n)}{Cap(V_{{\bf K}})} \leq \frac{n!}{n^n} O(1)$ with high
probability. This is true for the permanents of random matrices with nonnegative entries.

\item The most important question is whether or not there exists a FPRAS algorithm for the mixed volume (or for the mixed discriminant).
We conjecture that the answer is negative.
\item Another important open problem is whether or not
our mixed volume
generalization (\ref{schr ineq}) of Schrijver's lower bound on the number of perfect matchings in regular bipartite
graphs \cite{schr}, \cite{vor}, \cite{ejc-2008} is asymptotically sharp.

\end{enumerate}

\section{Acknowledgements}
The author is indebted to the both anonymous reviewers for a very careful and thoughtful
reading of the original (too long and undisciplined) version of this paper. Their numerous corrections
and suggestions are reflected in the current version.\\
I would like to thank the U.S. DOE for financial support through
Los Alamos National Laboratory's {\bf LDRD} program.

\appendix
\section{Proofs of Theorems (\ref{VDW}) and (\ref{S-V-G})}
\subsection{Useful (and well known) facts}
\fac \label{perm}
Let $\pi \in S_n$ be a permutation and  ${\bf K} = (K_1, \ldots, K_n)$ be a $n$-tuple of convex compact sets in $R^n$.
Then the next identity holds:
\beqn \label{perm-inv}
V(K_1, \ldots, K_n) = V(K_{\pi(1)},...,K_{\pi(n)})
\eeqn
\efac

\fac \label{AFin}
We recall here the fundamental Alexandrov-Fenchel inequalities for the mixed volume of $n$ convex sets in $R^n$:
\beqn \label{AF}
V(K_1,K_2,K_3,...,K_n)^2 \geq V(K_1,K_1,K_3,...,K_n)V(K_2,K_2,K_3,...,K_n)
\eeqn
\efac
\fac \label{twosets}
Let $(K_1,..,K_i), i < n-1$; $S,T$ be convex compact sets in $R^n$.
Define 
$$
a_0 = V(K_1,..,K_i,S,S,...,S), a_1 = V(K_1,..,K_i,S,S,...,S,T), a_{n-i} = V(K_1,..,K_i,T,T,...,T).
$$
Then the univariate polynomial $U$ defined by $U(t) = V(K_1,..,K_i,S+tT,S + tT,...,S + tT)$ is expressed as
\beqn
U(t) = \sum_{0 \leq j \leq n-i} {n-i \choose j} a_j t^j .
\eeqn
It follows from Facts (\ref{perm}) and the  Alexandrov-Fenchel inequalities that this univariate polynomial $U$ is
$(n-i)${\em -Newton}.
\efac

\fac \label{p-der}
Let ${\bf K} = (K_1, \ldots, K_n)$ be a $n$-tuple of convex compact sets in $R^n$. For a nonnegative
vector $(x_1,...,x_n)$, define the convex compact subset $K = \sum_{1 \leq i \leq n} x_i K_i$.
Then the following identity holds:
\beqn \label{segod}
\frac{\partial^{i}}{\partial x_{1}...\partial x_i} V_{{\bf K}}(x_1,...,x_n) = \frac{1}{(n-i)!}V(K_1,...,K_i,K,...,K)\;.
\eeqn
\efac

\fac \label{linr}
If $A$ is $n \times n$ real matrix and $\lambda_i > 0, 1 \leq i \leq n$ then
\beqn
V(\lambda_1 (A K_1),...,\lambda_n (A K_n)) = |\det(A)| (\prod_{1 \leq i \leq n} \lambda_i) V(K_1, \ldots, K_n)\;. 
\eeqn

\efac 

\fac \label{addit}.
Let $S,T;K_2,...,K_n$ be convex compact sets in $R^n$. The next (additivity) identity holds:
\beqn
V(S+T,K_2,...,K_n) = V(S,K_2,...,K_n) + V(T,K_2,...,K_n)
\eeqn
\efac

\fac \label{deg-af} Let ${\bf K} = (K_1, \ldots, K_n)$ be a $n$-tuple
of convex compact sets in $R^n$. Then the degree 
of the variable $x_i$ in the polynomial $V_ {{\bf K}}$ is
$deg_{V_ {{\bf K}}}(i) = aff(i)$, where $aff(i)$ is the affine dimension of $K_i, 1 \leq i \leq n$.  \efac

\subsection{Auxiliary univariate inequality}

\lem \label{aux}

Let $R(t) = \sum_{0 \leq i \leq k} a_i t^i$ be an $n-Newton$ polynomial, $n \geq k$ and
$$
\lambda(n,k) = \left(\min_{x > 0} \left(\frac{sv_{n,k}(x)}{x} \right)\right)^{-1},
$$
where the polynomial $sv_{n,k}(x) = 1+\sum_{1 \leq i \leq k} (\frac{x}{n})^{i} {n \choose i}$.

Then the following inequality holds:
\beqn \label{main-ine}
a_1 = R^{'}(0) \geq \lambda(n,k) \inf_{t > 0}\frac{R(t)}{t}
\;. \eeqn
Equality in (\ref{main-ine}) is attained if and only if $R(t) = R(0)\left(1+\sum_{1 \leq i \leq k} (\frac{a t}{n})^{i} {n \choose i} \right)$.

If $n=k$ then 
\beqn \label{facr}
\lambda(n,n) = \left(\min_{x > 0} \left(\frac{(1 + \frac{x}{n})^n}{x} \right) \right)^{-1} = \left(\frac{n-1}{n} \right)^{n-1} =: g(n).
\eeqn

Equality in (\ref{facr}) is attained iff $R(t) = R(0) \left(1 + \frac{at}{n} \right)^n$.
\elem

\prf
Note that $\lambda(n,k) \leq 1, k \geq 1$. If $R(0) = 0$ then clearly
$$
R^{'}(0) = \inf_{t > 0}\frac{R(t)}{t} \geq \lambda(n,k) \inf_{t > 0}\frac{R(t)}{t}.
$$
Therefore we need to consider only the case $R(0) > 0$. Assume WLOG that
$R(0) = 1$. It follows directly from the Newton inequalities (\ref{n-newton}) that
$$
a_i \leq {n \choose i} \left(\frac{a_{1}}{n} \right)^i, 1 \leq i \leq k.
$$
Thus
$$
R(x) = \sum_{0 \leq i \leq k} a_i x^i \leq 1 + \sum_{1 \leq i \leq k} \left(\frac{a_{1} x}{n} \right)^{i} {n \choose i} = sv(a_{1} x), x \geq 0.
$$
It also follows that
$$
\inf_{t > 0}\frac{R(t)}{t} \leq \min_{x > 0} \left(\frac{sv_{n,k}(a_{1} x)}{x} \right) = a_1 \min_{x > 0} \left(\frac{sv_{n,k}(x)}{x} \right),
$$
which gives that 
$$
R^{'}(0) = a_1 \geq \lambda(n,k) \inf_{t > 0}\frac{R(t)}{t}.
$$  
The remaining statements can be now easily verified.
\eprf
\cor \label{blad}
Denote as $Hom_{+}(n,n)$ a convex cone of homogeneous polynomials with nonnegative coefficients of degree $n$ in $n$ variables.
Consider polynomials 
$$
p \in Hom_{+}(n,n);~ q \in Hom_{+}(n-1,n-1), ~q(x_1,...,x_{n-1}) =\frac{\partial}{\partial x_{n}} p(x_1,...,x_{n-1},0). 
$$
Suppose that for all positive vectors $X=(x_1,...,x_{n-1})$ the univariate polynomials $R_{X}(t) = p(x_1,...,x_{n-1},t)$
are $n$-{\em Newton}.  Then the following inequality holds:
\beqn \label{caps}
Cap(q) \geq \lambda \left(n,deg_{p}(n) \right) Cap(p)
\;. \eeqn
\ecor
\prf
Note that because the coefficients of $p$ are nonnegative the (univariate) degree\\
 $deg(R_{X}) = deg_{p}(n)$ for all positive
vectors $X \in R_{++}^{n-1}$. It follows from the definition of the polynomial $q \in Hom_{+}(n-1,n-1)$ that
$$
q(x_1,...,x_{n-1}) = R_{X}^{'}(0).
$$
It follows from the definition (\ref{fir-cap}) of the {\bf capacity} that
$$
R_{X}(t) = p(x_1,...,x_{n-1},t) \geq Cap(p) (\prod_{1 \leq i \leq n-1} x_i) t.
$$
It now follows from the inequality (\ref{main-ine}) that
$$
q(x_1,...,x_{n-1}) = R_{X}^{'}(0) \geq \lambda \left(n,deg_{p}(n) \right) Cap(p) \prod_{1 \leq i \leq n-1} x_i,
$$
which gives that $Cap(q) \geq \lambda \left(n,deg_{p}(n) \right) Cap(p)$.
\eprf 
\subsection{Proof of Theorem (\ref{S-V-G})}
\prf
Let ${\bf K} = (K_1, \ldots, K_n)$ be a $n$-tuple of convex compact sets in $R^n$. 
 
We associate with the Minkowski polynomial
$$
V_{{\bf K}} \in Hom_{+}(n,n), V_{{\bf K}}(x_1,...,x_n) = V_{n}(x_1 K_1 +...+x_n K_n)
$$
the following sequence of
polynomials:

$q_n = V_{{\bf K}}; q_i \in Hom_{+}(i,i), q_i(x_1,...,x_i) = \frac{\partial^{n-i}}{\partial x_{i+1}...\partial x_n}q_{n}(x_1,...,x_i,0,...,0), 1 \leq i \leq n-1$. 

Note that 
$$
q_{1}(x) = x \frac{\partial^{n}}{\partial x_{1}...\partial x_n}V_{{\bf K}}(0,...,0) = V(K_1, \ldots, K_n) x,
$$
and 
$$
q_i(x_1,...,x_i) = \frac{\partial}{\partial x_{i+1}}q_{i+1}(x_1,...,x_i,0), 1 \leq i \leq n-1.
$$
Note also the obvious (but useful) inequality
\beqn \label{der-deg}
deg_{q_{i}}(i) \leq \min(i, deg_{q_{n}}(i)) = \min(i, aff(i))
\;. \eeqn
It follows from Fact(\ref{p-der}) and Fact(\ref{twosets}) that the polynomials $q_i, 2 \leq i \leq n$ satisfy the conditions
of Corollary(\ref{blad}). Therefore
\beqn \label{indu}
Cap(q_i) \geq \lambda \left(i+1,deg_{q_{i+1}}(i+1) \right) Cap(q_{i+1}) \geq \lambda \left(i+1,\min (i+1, aff(K_{i+1}) \right) Cap(q_{i+1}).
\eeqn

(We use here the inequality (\ref{der-deg}) and the fact that $\lambda(n,k)$ is strictly decreasing in both variables.)

Multiplying the inequalities (\ref{indu}), we get the inequality (\ref{main result}).

\eprf

\subsection{Proof of Theorem (\ref{VDW})}
\prf
{\bf (Proof of inequality (\ref{waer ineq}))}.\\
Sinse 
$$
\lambda(i,k) \geq \lambda(i,i) = g(i) =:\left(\frac{i-1}{i} \right)^{i-1}, 1 \leq k \leq i
$$
hence we get from (\ref{main result}) that
$$
V(K_1, \ldots, K_n) \geq Cap(V_{{\bf K}}) \prod_{2 \leq i \leq n} g(i) = \frac{n!}{n^n} Cap(V_{{\bf K}}).
$$
\eprf

\prf
{\bf (Proof of the uniqueness part of Theorem (\ref{VDW}))}.\\
As remarked above, we follow in the present paper the proof of uniqueness in \cite{ejc-2008}.
 
\begin{enumerate}
\item Assume that $Cap(V_{{\bf K}}) > 0$. Suppose that $l = aff(n) < n$. As $g(l) > g(n)$ hence 
$$
V(K_1, \ldots, K_n) \geq Cap(V_{{\bf K}}) g(l) \prod_{2 \leq i \leq n-1} g(i) > \frac{n!}{n^n} Cap(V_{{\bf K}}).
$$
Therefore if 
$$
V(K_1, \ldots, K_n) = \frac{n!}{n^n} Cap(V_{{\bf K}})
$$
then $aff(n) = n$. Using the permutation invariance (\ref{perm-inv}), we get that 
$$
aff(i) = n, 1 \leq i \leq n.
$$
In other words the convex compact sets $K_i$ all have nonempty interior.  
This fact together and the monotonicity of the mixed volume imply that all coefficients in the Minkowski polynomial $V_{{\bf K}}$ are strictly positive.
\item {\bf Scaling}.\\
All coefficients in the Minkowski polynomial $V_{{\bf K}}$ are strictly positive, hence there exists an unique positive vector
$(a_1,...,a_n)$ such that the scaled polynomial $p = V_{\{a_1K_1,...,a_nK_n\}}$ is doubly stochastic (see \cite{ejc-2008}):
$$
\frac{\partial}{\partial x_i} p(1,1,...,1) = 1, 1 \leq i \leq n. 
$$
We will deal, without loss of generality,
only with this doubly stochastic case.
\item {\bf Brunn-Minkowski}.\\
Let $(z_1,...,z_{n-1})$ be the unique minimizer of the problem 
$$
\min_{x_i > 0, 1 \leq i \leq n-1; \prod_{1 \leq i \leq n-1} x_i = 1} q_{n-1}(x_1,...,x_{n-1}). 
$$
Such an unique minimizer
exists as all the coefficients of $q_{n-1}$ are positive. 
It follows from Lemma (\ref{aux}) and the proof of Lemma(\ref{blad}), that 
$$
V_{{\bf K}}(z_1,...,z_{n-1},t) = V( S + tK_n) = (at +b)^{n}, S = \sum_{1 \leq i \leq n-1} z_i K_i
$$ 
for some positive numbers $a,b$. It follows from the equality case of the Brunn-Minkowski inequality \cite{BZ88} that
$$
K_n = \alpha S + \{T_n\}, \alpha > 0, T_n \in R^n.
$$
In other words,$K_n = \sum_{1 \leq j \leq n-1} A(n,j) K_j + \{T_n\}$, where $A_{n,j} > 0, 1 \leq j \leq n-1$ and $T_n \in R^n$.

In the same way, we get that
there exist a $n \times n$ matrix $A$, with the zero diagonal and positive off-diagonal part, and vectors $T_1,...,T_n \in R^n$
such that 
$$
K_i = \sum_{j \neq i} A(i,j) K_j + \{T_i\}. 
$$
It follows from the doubly-stochasticity of the polynomial $V_{{\bf K}}$ that all row sums
of the matrix $A$ are equal to one. Indeed, using the identity (\ref{segod}), we get that
$$
(n-1)! = (n-1)! \frac{\partial}{\partial x_i} V_{{\bf K}}(1,1,...,1) = V(SUM,SUM,...,SUM,K_i); SUM = K_1+...+K_n, 1 \leq i \leq n.
$$
As $K_i = \sum_{j \neq i} A(i,j) K_j + \{T_i\}$, we get, using Fact(\ref{addit}) and Fact(\ref{linr}), that
$$
(n-1)! =  V(SUM,SUM,...,SUM,\sum_{j \neq i} A(i,j) K_j + \{T_i\}) = 
$$

$$
= \sum_{j \neq i} A(i,j)V(SUM,SUM,...,SUM,K_j) = (n-1)!\sum_{j \neq i} A(i,j).
$$

Therefore $\sum_{j \neq i} A(i,j) = 1, 1 \leq i \leq n$.

\item Associate with the convex compact set $K_i \subset R^n$ its support function 
$$
\gamma_{i}(X) = \max_{Y \in K_{i}} <X,Y>, X \in R^n.
$$
We get that 
$$
\gamma_{i}(X) = \sum_{j \neq i} A(i,j)\gamma_{j}(X)  + <X, T_{i}>, X \in R^n.
$$
As the kernel
$$
Ker(I-A) = \{Y \in R^n : (I-A)Y = 0\} = \{c (1,1,...,1), c \in R \},
$$
it follows finally that
$$
\gamma_{i}(X) = \alpha(X) + <X, L_j>, X \in R^n
$$
for some functional $\alpha(X)$ and vectors $L_1,...,L_n \in R^n$.  This 
means, in the doubly-stochastic case, that $K_i = K_1 + \{L_i - L_1 \}, 2 \leq i \leq n$. 
\end{enumerate}
\eprf

\section{Inequalities for Minkowski and Minkowski-like Polynomials}
Let $f : R^n_{++} \rightarrow R_{++}$ be a differentiable positive-valued functional defined on the strictly positive
orthant $R^n_{++}$. We assume that $f$ is $n$-homogeneous, i.e. that $f(a x_1,...,a x_n) = a^n f(x_1,...,x_n)$ and
the partial derivatives $\frac{\partial}{\partial x_i} f(x_1,...,x_n) > 0 :(x_1,...,x_n) \in R^n_{++}$. 
We denote the set of such homogeneous functionals as $PoH(n)$.

We define the {\bf capacity} as
$$
Cap(f) = \inf_{x_i > 0} \frac{f(x_1,...,x_n)}{\prod_{1 \leq i \leq n} x_i } = 
\inf_{x_i > 0,\prod_{1 \leq i \leq n} x_i =1 }f(x_1,...,x_n).
$$
We define two subsets of $PoH(n)$:\\
$Cav(n)$ - consisting of $f \in PoH(n)$ such that $f^{\frac{1}{n}}$ is concave on all
half-lines $\{X + t Y : t \geq 0 \} : X,Y \in R^n_{++}$ ; $Vex(n)$ - consisting of $f \in PoH(n)$ such 
that $f^{\frac{1}{n}}$ is convex on all
half-lines $\{X + t Y : t \geq 0 \} : X,Y \in R^n_{++}$.

Recall theBrunn-Minkowski theorem : the Minkowski polynomial $V_{{\bf K}}(x_1,...,x_n)$
 belongs to $Cav(n)$. Therefore the results in
this Appendix apply to the Minkowski polynomials.

We also define the following Generalized Sinkhorn Scaling :
$$
SH(x_1,...,x_n) = (y_1,...,y_n) : y_i = \frac{f(x_1,...,x_n)}{\frac{\partial}{\partial x_i} f(x_1,...,x_n)} = \frac{x_i}{\gamma_i}, 
\gamma_i = \frac{x_i \frac{\partial}{\partial x_i} f(x_1,...,x_n)} {f(x_1,...,x_n)}.
$$

\thm \label{Sink}
If $f \in Cav(n)$ then the following inequality holds:
\beqn \label{main}
f(SH(x_1,...,x_n)) \leq f(x_1,...,x_n)
\;. \eeqn
If $f \in Vex(n)$ then the reverse inequality holds:
\beqn
f(SH(x_1,...,x_n)) \geq f(x_1,...,x_n)
\;. \eeqn
\ethm

\prf
Let $X =(x_1,...,x_n) \in R^n_{++}$ and $Y = SH(x_1,...,x_n)$. We can assume without loss of generality
that $f(X) = 1$. If $f \in Cav(n)$ then the univariate function $g(t) = (f(X + tY))^{\frac{1}{n}}$
is concave for $t \geq 0$. Therefore
$$
g(t) \leq (g(0) + \frac{g^{\prime}(0)}{n} t)^{n} = (1 + \frac{g^{\prime}(0)}{n} t)^{n}.
$$
We get, by elementary calculus, that 
$$
g^{\prime}(0) = \sum_{1 \leq i \leq n}(\frac{\partial}{\partial x_i} f(x_1,...,x_n)) x_i = n.
$$
The functional $f$ is $n$-homogeneous, hence 
$g(t) = t^{n} f(Y + t^{-1} X) \leq (1 + t)^{n}$, and finally 
$f(Y + t^{-1} X) \leq (\frac{1+t}{t})^{n}$.  
Taking the limit $t \rightarrow \infty$ we get $f(SH(x_1,...,x_n)) \leq 1 = f(x_1,...,x_n)$.

The convex case is proven in the very same way.
\eprf

Theorem (\ref{Sink}) suggests the following algorithm to approximate $Cap(f)$:
\beqn \label{alg}
X_{n+1} = Nor(SH(X_{n})) : Nor(x_1,...,x_n) = (\frac{x_1}{a},...,\frac{x_n}{a}), a = \sqrt[n]{(\prod_{1 \leq i \leq n} x_i)}.
\eeqn

\cor \label{nemir}
Consider $f \in Cav(n)$. Suppose that $Cap(f) > 0$,
$$
\log(Cap(f)) \leq \log(f(x_1,...,x_n)) \leq \log(Cap(f)) + \epsilon, 0 < \epsilon \leq \frac{1}{10}
$$
and $\prod_{1 \leq i \leq n} x_i = 1; x_i > 0, 1 \leq i \leq n$. Then
\beqn
\sum_{1 \leq i \leq n} (1 - \frac{x_i \frac{\partial}{\partial x_i} f(x_1,...,x_n)} {f(x_1,...,x_n)})^2 \leq 10
\epsilon
\;. \eeqn
\ecor
\prf
Let $\gamma_i = \frac{x_i \frac{\partial}{\partial x_i} f(x_1,...,x_n)} {f(x_1,...,x_n)}$. It follows
from the Euler's identity that $\sum_{1 \leq i \leq n} \gamma_i = n$ and
thus $\log(\prod_{1 \leq i \leq n} \gamma_i) \leq 0$.

Inequality (\ref{main}) can be rewritten
as
$$
f(\frac{x_{1}}{\gamma_{1}},...,\frac{x_{n}}{\gamma_{n}} ) \leq f(x_1,...,x_n).
$$
Therefore $\log(Cap(f)) \leq Cap(f) + \epsilon + \log(\prod_{1 \leq i \leq n} \gamma_i)$,
which gives the inequality 
$$
-\epsilon \leq \log(\prod_{1 \leq i \leq n} \gamma_i) \leq 0.
$$
Finally, using Lemma 3.10 in \cite{LSW-2000}, we see that $\sum_{1 \leq i \leq n}(1 - \gamma_i)^2 \leq 10 \epsilon$.
\eprf 

Corollary (\ref{nemir}) generalizes (with a much more transparent proof) corresponding results from \cite{GS1} and \cite{nr}.\\

{\it The following Lemma proves inequality (\ref{sec-der}).}

\lem \label{vtor}
\begin{enumerate}
\item
Let $p(t) = \sum_{0\leq i \leq n} a_i t^i, a_i \geq 0$ be a polynomial with nonnegative coefficients.
Assume that $\log(p(t))$ is concave on $R_{++}$ and define $q(x) = \log(p(e^{x}))$. Then
$q(x)$ is convex on $R$ and its second derivative satisfies the following inequality
\beqn \label{log}
0 \leq q^{\prime \prime}(x) \leq n
\;. \eeqn
\item
Let $p(t) = \sum_{0\leq i \leq n} a_i t^i, a_i \geq 0$ be a polynomial with nonnegative coefficients.
Assume that $p(t)^{\frac{1}{m}}, m \geq n$ is concave on $R_{+}$. Then
\beqn \label{sec-bnd}
0 \leq q^{\prime \prime}(x) \leq f(n,m),
\eeqn
where $f(n,m) = n - \frac{n^2}{m}$ if $n \leq \frac{m}{2}$; and $f(n,m) = \frac{m}{4}$ otherwise.
If $n=m$ then $f(n,m) = \frac{n}{4}$ and the upper bound (\ref{sec-bnd}) is attained
on polynomials $p(t) = (a + tb)^n ; a,b > 0$.
\end{enumerate}

\elem

\prf
\begin{enumerate}
\item
The convexity of $q(x)$ is well known. The concavity of $\log(p(t))$ is equivalent to the
inequality $(p^{\prime}(t))^2 \geq p(t)p^{\prime \prime}(t) : t \geq 0$. 
Putting $y =e^{x}$, we get that
$$
q^{\prime \prime}(x) = \frac{p^{\prime \prime}(y) y^2}{p(y)} + \frac{p^{\prime}(y) y}{p(y)} - \left(\frac{p^{\prime}(y) y}{p(y)} \right)^2
$$
The concavity of $\log(p(t))$ gives that $\frac{p^{\prime \prime}(y) y^2}{p(y)} - \left(\frac{p^{\prime}(y) y}{p(y)} \right)^2 \leq 0$.

The coefficients of the polynomial $p$ are nonnegative, hence $\frac{p^{\prime}(y) y}{p(y)} \leq n$. This last observation
proves that 
$$
q^{\prime \prime}(x) \leq \frac{p^{\prime}(y) y}{p(y)} \leq n.
$$
\item
Our proof of (\ref{sec-bnd}) is a direct adaptation of the above proof of (\ref{log}).  
We use the following characterization
of the concavity of $p(t)^{\frac{1}{m}}, m \geq n$: \\
$(p^{\prime}(t))^2 \geq \frac{m}{m-1} p(t) p^{\prime \prime}(t) : t \geq 0$.
\end{enumerate}

\eprf

We note that just the nonnegativity of the coefficients implies the quadratic bound 
$$
q^{\prime \prime}(x) \leq  0.25 n^2.
$$

\end{document}